\documentclass{article}

\usepackage{diagrams}
\usepackage{latexsym}
\usepackage{amsfonts}
\def\text#1{\mbox{\rm #1\ }}

\def\ie{{\rm i.e.,\/}\ }
\def\etc{{\rm etc.\/}\ }
\def\cf{{\rm cf.\/}\ }
\def\id{\mbox{\it id\,}}
\def\one{\mbox{\rm 1}\hskip-2.8pt \mbox{\rm l}}
\newcommand{\ZZ}{\mathbb{Z}}
\newcommand{\RR}{\mathbb{R}}
\newcommand{\CC}{\mathbb{C}}
\def\cro#1{[ {#1} ]}
\def\Mlo#1{\left( M_{#1}(\Lambda^2)\right)_o}
\newcommand{\inplus}{\ensuremath{\,\,\subset\!\!\!\!\!\!+\,\,\,\,}}
\title{Geometry of the reduced quantum plane\thanks{~Contribution to the
       proceedings of the conference ``Quantum Groups and Fundamental
       Physical Applications'', I.S.I. Guccia, Palerme, December 1997}
   \vspace{0.7cm} \\
}

\author{R. Coquereaux${}^1$\thanks{~Email: coque@cpt.univ-mrs.fr},
        A. O. Garc\'{\i}a${}^2$\thanks{~Email: ariel@cab.cnea.edu.ar},
        R. Trinchero${}^2$\thanks{~Email: trincher@cab.cnea.edu.ar} \\
\\
${}^1$ {\it Centre de Physique Th\'eorique - CNRS - Luminy, Case 907} \\
       {\it F-13288 Marseille Cedex 9 - France} \\
\\
${}^2$ {\it Instituto Balseiro and Centro At\'omico Bariloche} \\
       {\it CC 439 - 8400 - Bariloche - R\'{\i}o Negro - Argentina} \\
\\
}

\date{October 10, 1998}

\begin{document}

\thispagestyle{empty}
\begin{titlepage}

\maketitle

\vfill

\abstract{
We consider the space $\mathcal M$ of $N\times N$ matrices as a reduced
quantum plane and discuss its geometry under the action and coaction of
finite dimensional quantum groups (a quotient of $U_qSL(2)$, $q$ being an
$N$-th root of unity, and its dual). We also introduce a differential
calculus for $\mathcal M$: a quotient of the Wess Zumino complex. We
shall restrict ourselves to the case $N$ odd and often choose the
particular value $N=3$.

The present paper (to appear in the proceedings of the conference
``Quantum Groups and Fundamental Physical Applications'', Palerme,
December 1997) is essentially a short version of {\tt math-ph/9807012}.
}

\vspace{1.2 cm}


\vspace{1.0cm}

\noindent Anonymous ftp or gopher: cpt.univ-mrs.fr

\vspace{0.5 cm}

\noindent {\tt math-ph/9811017} \\
\noindent CPT-98/P.3704 \\

\vspace*{0.3 cm}

\end{titlepage}

%%%%%%%%%%%%%%%%%%%%%%%%%%%%%%%%%%%%%%%%%%%%%%%%%%%%%%%%%%%%%%%%%%%%%%%%%%
%%%%%%%%%%%%%%%%%%%%%%%%%%%%%%%%%%%%%%%%%%%%%%%%%%%%%%%%%%%%%%%%%%%%%%%%%%

\section{Introduction}

When $q$ is a root of unity ($q^{N}=1$), the quantized enveloping algebra
$U_{q}sl(2,\CC)$ possesses interesting quotients that are finite dimensional
Hopf algebras. The structure of the left regular representation of such an
algebra was investigated in \cite{Alekseev} and the pairing with its dual in
\cite{Gluschenkov}. We call ${\mathcal H}$ the Hopf algebra quotient of
$U_{q}sl(2,\CC)$ defined by the relations $K^N=\one$, $X_{\pm}^N=0$
(we shall define the generators $K, X_{\pm}$ in a later section), and
${\mathcal F}$ its dual. It was shown\footnote{
Warning: the authors of \cite{Alekseev} actually consider a Hopf algebra
quotient defined by $K^{2N} = 1$, $X_{\pm}^N=0$, so that their algebra is,
in a sense, twice bigger than ours.
}
in \cite{Alekseev} that the {\underline{non}} semi-simple algebra
${\mathcal H}$ is isomorphic with the direct sum of a complex matrix
algebra and of several copies of suitably defined matrix algebras with
coefficients in the ring $Gr(2)$ of Grassmann numbers with two generators.
The explicit structure (for all values of $N$) of those algebras, including
the expression of generators themselves, in terms of matrices with
coefficients in $\CC$ or $Gr(2)$ was obtained by \cite{Ogievetsky}. Using
these results, the representation theory of ${\mathcal H}$, for the case
$N=3$, was presented in \cite{Coquereaux}. Following this work, the authors
of \cite{Dabrowski} studied the action of ${\mathcal H}$ (case $N=3$) on the
algebra of complex matrices ${\mathcal M} \equiv M_3(\CC)$. In the letter
\cite{CoGaTr} a reduced Wess-Zumino complex $\Omega_{WZ}({\mathcal M})$ was
introduced, thus providing a differential calculus bicovariant with respect
to the action of the quantum group ${\mathcal H}$ on the algebra $M_3(\CC)$
of complex matrices. This differential algebra (that could be used to
generalize gauge field theory models on an auxiliary smooth manifold) was
also analysed in terms of representation theory of ${\mathcal H}$ in the
same work. In particular, it was shown that $M_3(\CC)$ itself can be
reduced into the direct sum of three indecomposable representations of
${\mathcal H}$. This result was generalized in \cite{Coquereaux-Schieber}.
A general discussion of several other properties of the dually paired Hopf
algebras ${\mathcal F}$ and ${\mathcal H}$ (scalar products, star
structures, twisted derivations, \etc) can also be found there, as well as
in the article \cite{CoGaTr-E}. In the present contribution we
present a summary of results already discussed in the papers
\cite{CoGaTr, CoGaTr-E, Coquereaux-Schieber}. The original purpose
of our work was to define generalized differential forms and
generalized connections on a smooth manifold, with values in the
Lie algebra of the linear group
$GL(N)$, in such a way that there would be a non trivial global
action of some Hopf algebra on the so obtained differential
complex, extending somehow the usual action of the isometry group
of the manifold (when it exists) by some internal quantum symmetry.
This construction will be recalled in the text.

%%%%%%%%%%%%%%%%%%%%%%%%%%%%%%%%%%%%%%%%%%%%%%%%%%%%%%%%%%%%%%%%%%%%%%%%%%

\section{The space $\mathcal M$ of $N \times N$ complex matrices as a
         reduced quantum plane}
\label{sec:red-q-plane}

The algebra of $N \times N$ matrices can be generated by two elements
$x$ and $y$ with relations:
\begin{equation}
   xy = q yx \qquad \text{and} \qquad x^N = y^N = \one \ ,
\label{M-relations}
\end{equation}
where $q$ denotes an $N$-th root of unity ($q \neq 1$) and $\one$ denotes
the unit matrix. Explicitly, $x$ and $y$ can be taken as the following
matrices:
\begin{equation}
x = \begin{pmatrix}
    { 1 & & & & \cr
       & q^{-1} & & & \cr
       & & q^{-2} & & \cr
       & & & \ddots & \cr
       & & & & q^{-(N-1)} }
\end{pmatrix}
\qquad
y = \begin{pmatrix}
    { 0 & & & \cr
   \vdots & & \one_{N-1} & \cr
      0 & & & \cr
      1 & 0 & \cdots & 0}
\end{pmatrix}
\label{xymatrix}
\end{equation}
This result can
be found in \cite{Weyl}. \\
{\bf Warning}: for technical reasons, we shall assume in all this paper
that $N$ is odd and that $q$ is a {\sl primitive} root of unity.

Here and below, we shall simply denote the algebra $M_{N}(\CC)$ by the
symbol ${\mathcal M}$, and call it ``the reduced quantum plane''.

%%%%%%%%%%%%%%%%%%%%%%%%%%%%%%%%%%%%%%%%%%%%%%%%%%%%%%%%%%%%%%%%%%%%%%%%%%

\section{The finite dimensional quantum group $\mathcal F$ and its
         coaction on $\mathcal M$}
\label{sec:q-group-F}

One introduces {\sl non-commuting symbols\/} $a,b,c$ and $d$ generating an
algebra ${\mathcal F}$ and imposes that the quantities $x',y'$ (and
$\tilde x, \tilde y$) obtained by the following matrix equalities should
satisfy the same relations as $x$ and $y$.
\begin{equation}
\Delta_L \pmatrix{x \cr y} = \pmatrix{a & b \cr c & d}
                             \otimes \pmatrix{x \cr y} \doteq
         \pmatrix{x' \cr y'} \qquad \text{left coaction} \ ,
\label{3.1}
\end{equation}
and
\begin{equation}
\Delta_R \pmatrix{x & y} = \pmatrix{x & y} \otimes
                           \pmatrix{a & b \cr c & d} \doteq
         \pmatrix{\tilde x & \tilde y} \qquad \text{right coaction} \ .
\label{3.2}
\end{equation}

These equations also define left and right coactions, that extend to the
whole of $\mathcal M$ using the homomorphism property
$\Delta_{L,R}(fg) = \Delta_{L,R}(f) \Delta_{L,R}(g) \;\;$
$(f,g \in {\mathcal M})$. Here one should not confuse $\Delta$ (the
coproduct on a quantum group that we shall introduce later) with
$\Delta_{R,L}$ (the $R,L$-coaction on $\mathcal M$)!

The elements $a,b,c,d$ should therefore satisfy an algebra such that
\begin{eqnarray}
   \Delta_L (xy - q yx) &=& 0  \\
   \Delta_L (x^{N} -1) = \Delta_L (y^{N} -1) &=& 0 \ ,
\end{eqnarray}
and the same for $\Delta_R$. This leads to the usual relations defining
the algebra of ``functions over the quantum plane'' \cite{Manin}:
\begin{equation}
\begin{array}{ll}
   qba = ab \qquad & qdb = bd \\
   qca = ac        & qdc = cd \\
   cb = bc         & ad-da = (q-q^{-1})bc \ , \\
\end{array}
\end{equation}
but, we also have non quadratic relations:
\begin{equation}
\begin{tabular}{ll}
   $a^N = \one \ , $ & $ b^N = 0    \ , $ \\
   $c^N = 0    \ , $ & $ d^N = \one \ . $
\end{tabular}
\label{F-products-quotient}
\end{equation}

The element ${\mathcal D} \doteq da -q^{-1}bc = ad - qbc $ is central
(it commutes with all the elements of $Fun(GL_q(2))$); it is called the
$q$-determinant and we set it equal to $\one$. Since $a^N = \one$,
multiplying the relation $ad = \one + qbc$ from the left by $a^{(N-1)}$
leads to
\begin{equation}
   d = a^{(N-1)} (\one + qbc) \ ,
\label{d-element-in-F}
\end{equation}
so $d$ is not needed and can be eliminated. The algebra $\mathcal F$ can
therefore be {\sl linearly \/} generated ---as a vector space--- by the
elements $a^\alpha b^\beta c^\gamma $ where indices $\alpha, \beta, \gamma$
run in the set $\{0,1,\ldots ,N-1\}$. We see that $\mathcal F$ is a
{\sl finite dimensional\/} associative algebra, whose dimension is
$$
   \dim ({\mathcal F}) = N^{3} \ .
$$

${\mathcal F}$ is not only an associative algebra but a Hopf algebra, with
the corresponding maps defined on the generators as follows:

\begin{description}
\item[Coproduct:]
   $\Delta a = a \otimes a + b \otimes c$,
   $\Delta b = a \otimes b + b \otimes d$,
   $\Delta c = c \otimes a + d \otimes c$,
   $\Delta d = c \otimes b + d \otimes d$.

\item[Antipode:] $Sa = d$, $Sb = -q^{-1} b$, $Sc = -q c$, $Sd = a$.

\item[Counit:] $\epsilon(a)=1$, $\epsilon(b)=0$, $\epsilon(c)=0$,
               $\epsilon(d)=1$.
\end{description}

We call $\mathcal F$ the {\sl reduced quantum unimodular group\/} associated
with an $N$-th root of unity. It is, by construction, an associative
algebra. However, it is not semi-simple. Therefore, $\mathcal F$ is not a
matrix quantum group in the sense of Woronowicz \cite{Woronowicz}.

The coaction of $\mathcal F$ on $\mathcal M$ was given above.
Actually, $\mathcal M$ endowed with the two coactions $\Delta_L$
and $\Delta_R$ is a left and right comodule {\em algebra} over $\mathcal F$,
\ie a corepresentation space of the quantum group $\mathcal F$ such that
\begin{eqnarray}
   \Delta_{L,R}(zw) &=& \Delta_{L,R}(z) \, \Delta_{L,R}(w) \nonumber \\
   \Delta_{L,R}(\one) &=& \one {\otimes} \one \ .
\label{comodule-algebra-condition}
\end{eqnarray}

%%%%%%%%%%%%%%%%%%%%%%%%%%%%%%%%%%%%%%%%%%%%%%%%%%%%%%%%%%%%%%%%%%%%%%%%%%
%%%%%%%%%%%%%%%%%%%%%%%%%%%%%%%%%%%%%%%%%%%%%%%%%%%%%%%%%%%%%%%%%%%%%%%%%%

\section{The dual $\mathcal H$ of $\mathcal F$, and its action
         on $\mathcal M$}
\label{sec:q-group-H}

Being $\mathcal F$ a quantum group (a Hopf algebra), its dual
${\mathcal H} \doteq {\mathcal F}^*$ is a quantum group as well. Let
$u_i \in \mathcal F$ and $X_i \in \mathcal H$. We call $< X_i, u_j >$
the evaluation of $X_i$ on $u_j$ (a complex number).

\begin{itemize}
\item
   Starting with the coproduct $\Delta$ on $\mathcal F$, one defines a
   product on $\mathcal H$, by
   $<X_1 X_2, u>\doteq <X_1 \otimes X_2, \Delta u >$.
\item
   Using the product in $\mathcal F$, one defines a coproduct (that we
   again denote $\Delta$) in $\mathcal H$ by
   $<\Delta X, u_1 \otimes u_2 > \doteq < X, u_1 u_2>$.
\item
   The interplay between unit and counit is given by:
   $<\one_{\mathcal H}, u> = \epsilon_{\mathcal F}(u)$
   and $< X,\one_{\mathcal F}> = \epsilon_{\mathcal H}(X)$.
\end{itemize}

\noindent
The two structures of algebra and coalgebra are clearly interchanged by
duality.

It is clear that $\mathcal H$ is a vector space of dimension $N^{3}$. It
can be generated, as a complex algebra, by elements $X_\pm$, $K$ dual to
the generators of $\mathcal F$:
$$
   <K,a> = q \qquad <X_+,b> = 1 \qquad <X_-,c> = 1 \ ,
$$
all other pairings between generators being zero. In this way we get:

\begin{description}

\item[Multiplication:]
\begin{eqnarray}
   K X_{\pm}     &=& q^{\pm 2} X_{\pm} K                 \nonumber \\
   \left[X_+ , X_- \right]
                 &=& {1 \over (q - q^{-1})} (K - K^{-1}) \\
   K^N           &=& \one                                \nonumber \\
   X_+^N = X_-^N &=& 0 \ .                               \nonumber
\nonumber
\label{H-products}
\end{eqnarray}

\item[Comultiplication:]
\begin{eqnarray}
   \Delta X_+ & = & X_+ \otimes \one + K \otimes X_+      \nonumber \\
   \Delta X_- & = & X_- \otimes K^{-1} + \one \otimes X_-           \\
   \Delta K   & = & K \otimes K                           \nonumber \\
   \Delta K^{-1} & = & K^{-1} \otimes K^{-1} \ .          \nonumber
\nonumber
\label{H-coproducts}
\end{eqnarray}
It extends to the whole $\mathcal H$ as an algebra morphism, \ie
$\Delta(XY) = \Delta X \, \Delta Y$.

\item[Antipode:]
It is defined by:
   $ S \one = \one $,
   $ S K = K^{-1} $,
   $ S K^{-1} = K $,
   $ S X_+ = - K^{-1} X_+ $,
   $ S X_- = - X_- K $,
and it extends as an anti-automorphism, \ie $S(XY) = SY \, SX$. As
usual, the square of the antipode is an automorphism, given by
$S^2 u = K^{-1} u K$.

\item[Counit:]
The counit $\epsilon$ is defined by
   $ \epsilon \one = \epsilon K = \epsilon K^{-1} = 1 $,
   $ \epsilon X_+ = \epsilon X_- = 0 $.

\end{description}

Warning: When $q^N=1$, one can also factorize the universal algebra over the
relations $K^{2N} = \one$, $X_{\pm}^N = 0$, rather than $K^N = \one$,
$X_{\pm}^N = 0$. These relations also define a Hopf ideal but the obtained
Hopf algebra is twice as big as ours ($K^{N}$ is then a central element
but is not equal to $\one$).

%%%%%%%%%%%%%%%%%%%%%%%%%%%%%%%%%%%%%%%%%%%%%%%%%%%%%%%%%%%%%%%%%%%%%%%%%%

\subsection{Action of $\mathcal H$ on $\mathcal M$}
\label{subsec:actions-of-H}

Using the fact that $\mathcal F$ coacts on $\mathcal M$ in two possible
ways, and that elements of $\mathcal H$ can be interpreted as distributions
on $\mathcal F$, we obtain two commuting actions of $\mathcal H$ on the
quantum space $\mathcal M$. We shall describe the left action for arbitrary
elements and give explicit results for the generators.

Let $z \in \mathcal M$, $X \in \mathcal H$, and
$\Delta_R z = z_\alpha \otimes v_\alpha$ with $z_\alpha \in \mathcal M$
and $v_\alpha \in \mathcal F$ (implied summation). The operation
\begin{equation}
   X^L[z] \doteq (\id \otimes <X,\cdot>) \Delta_R z
               = <X, v_\alpha> z_\alpha \ .
\label{L-H-action-on-M}
\end{equation}
is a {\em left} action of $\mathcal H$ on $\mathcal M$ (dual to the
{\em right}-coaction of $\mathcal F$). With this $L$-action we can check
that $\mathcal M$ is indeed a left-$\mathcal H$-module algebra.

For the case $N=3$, complete tables are given in \cite{CoGaTr, CoGaTr-E},
and ---with other conventions--- in \cite{Dabrowski}. The results can be
summarized as follows:
\begin{eqnarray}
K^L \cro{x^{r}y^{s}}   &=& q^{(r-s)}x^{r}y^{s}                   \nonumber \\
X_+^L \cro{x^{r}y^{s}} &=& q^{r}(\frac{1-q^{-2s}}{1-q^{-2}})x^{r+1}y^{s-1} \\
X_-^L \cro{x^{r}y^{s}} &=& q^{s}(\frac{1-q^{-2r}}{1-q^{-2}})x^{r-1}y^{s+1}
   \nonumber
\label{actionofHonM}
\end{eqnarray}
with $1 \le r,s \le N$.

As ${\mathcal M}$ is a module (a representation space) for the quantum group
${\mathcal H}$, we can reduce it into indecomposable modules. To this end it
is necessary to know at least part of the representation theory of
${\mathcal H}$, but note that for the $N=3$ case we have (up to
multiplicative factors)

\vspace{0.5cm}
\begin{equation}
\begin{diagram}
  y^2           & \rDotsto 0  \\
  \uDotsto{X_-^L} \dTo{X_+^L} \\
  xy            &             \\
  \uDotsto{X_-^L} \dTo{X_+^L} \\
  x^2           & \rTo 0
\end{diagram}
\quad\hbox{,}\quad
\begin{diagram}
  y             & \rDotsto  & 0       \\
                & \luTo     &         \\
\uDotsto{X_-^L} \dTo{X_+^L}
                &           & x^2 y^2 \\
                & \ldDotsto &         \\
  x             & \rTo      & 0
\end{diagram}
\quad\hbox{and}\quad
\begin{diagram}
  xy^2          &           &               \\
                & \rdDotsto &               \\
\uDotsto{X_-^L} \dTo{X_+^L}
                &           & \one          \\
                & \ruTo     & \dDotsto \dTo \\
  x^2 y         &           & 0
\end{diagram}
\label{graph:left-H-action-on-M}
\end{equation}
\vspace{0.3cm}

\noindent We see clearly on these diagrams that the algebra of $3\times 3$
matrices can be written as a sum of three inequivalent, 3-dimensional,
indecomposable representations of ${\mathcal H}$: an irreducible one, and
two indecomposable but reducible modules (each of these two contains a non
trivial invariant subspace).

%%%%%%%%%%%%%%%%%%%%%%%%%%%%%%%%%%%%%%%%%%%%%%%%%%%%%%%%%%%%%%%%%%%%%%%%%%

\subsection{The structure of the non semisimple algebra $\mathcal H$}
\label{subsec:structure-of-H}

Using a result by \cite{Alekseev}, the explicit structure (for all values
of $N$) of those algebras, including the expression of generators
$X_{\pm}, K$ themselves, in terms of matrices with coefficients both in
$\CC$ and in the Grassmann\footnote{
Remember that $\theta_1^{2} = \theta_2^{2} = 0$ and that
$\theta_1 \theta_2 = -\theta_2 \theta_1$.
}
algebra $Gr(2)$ with two generators $\theta_1,\theta_2$, was obtained by
\cite{Ogievetsky} (it was explicitly described for $N = 3$ by
\cite{Coquereaux}, see also \cite{Coquereaux-Schieber} for a general $N$).

We shall not need the general theory but only the following fact:
when $N$ is odd, ${\mathcal H}$ is isomorphic with the direct sum
\begin{equation}
   \mathcal{H} = M_N \oplus \Mlo{N-1|1} \oplus \Mlo{N-2|2} \oplus \cdots
                     \cdots \oplus \Mlo{\frac{N+1}{2}|\frac{N-1}{2}}
\label{isomorphism}
\end{equation}
where:
\begin{itemize}

\item[-] $M_N$ is an $N\times N$ complex matrix

\item[-] An element of the $\Mlo{N-p|p}$ part (space that we shall just
   call $M_{N-p|p}$) is an $(N-p,p)$ block matrix of the following form:
   \begin{equation}
      \begin{pmatrix}{
         \bullet & \cdots & \bullet & \circ  & \cdots & \circ  \cr
         \vdots  & \scriptscriptstyle{(N-p)\times(N-p)}
                          & \vdots  & \vdots &        & \vdots \cr
         \bullet & \cdots & \bullet & \circ  & \cdots & \circ  \cr
         \circ   & \cdots & \circ  & \bullet & \cdots & \bullet \cr
         \vdots  &        & \vdots  & \vdots &
                        \scriptscriptstyle{p\times p} & \vdots \cr
         \circ   & \cdots & \circ  & \bullet & \cdots & \bullet }
      \end{pmatrix}
   \end{equation}
   We have introduced the following notation: \\
   $\bullet$ is an even element of the ring $Gr(2)$ of Grassmann numbers
   with two generators, \ie of the kind
   $\bullet = \alpha + \beta \theta_1 \theta_2$, $\alpha,\beta \in \CC$. \\
   $\circ$ is an odd element of the ring $Gr(2)$ of Grassmann numbers with
   two generators, \ie $\circ = \gamma\theta_1 + \delta \theta_2$,
   $\gamma, \delta \in \CC$.
\end{itemize}
When $N$ is even, the discussion depends upon the parity of $N/2$ and
we shall not discuss this here.

Notice that ${\mathcal H}$ is \underline{not} a semi-simple algebra:
its Jacobson radical ${\mathcal J}$ is obtained by selecting in equation
(\ref{isomorphism}) the matrices with elements proportional to Grassmann
variables. The quotient ${\mathcal H} / {\mathcal J}$ is then
semi-simple\ldots but no longer Hopf!

Projective indecomposable modules (PIM's, also called principal modules) for
${\mathcal H}$ are directly given by the columns of the previous matrices.
\begin{itemize}
   \item[-] From the $M_{N}$ block, one obtains $N$ equivalent irreducible
      representations of dimension $N$ that we shall denote $N_{irr}$.
   \item[-] From the $M_{{N-p}\vert p}$ block (\underline{assume} $p<N-p$),
      one obtains
      \begin{itemize}
         \item $(N-p)$ equivalent indecomposable projective modules of
            dimension $2N$ that we shall denote $P_{N-p}$ with
            elements of the kind
            $$
            (\underbrace{\bullet \bullet \cdots \bullet}_{N-p}
            \underbrace{\circ \circ \cdots \circ}_{p})
            $$
         \item $p$ equivalent indecomposable projective modules (also
            of dimension $2N$) that we shall denote $P_{p}$ with
            elements of the kind
            $$
            (\underbrace{\circ \circ \cdots \circ}_{N-p}
            \underbrace{\bullet \bullet \cdots \bullet}_{p})
            $$
      \end{itemize}
\end{itemize}

Other submodules can be found by restricting the range of parameters
appearing in the columns defining the PIM's and imposing stability under
multiplication by elements of ${\mathcal H}$. In this way one can
determine, for each PIM, the lattice of its submodules. For a given PIM of
dimension $2N$ (with the exception of $N_{irr}$), one finds totally ordered
sublattices (displayed below) with exactly three non trivial terms: the
radical (here, it is the biggest non trivial submodule of a given PIM),
the socle (here it is the smallest non trivial submodule), and one
``intermediate'' submodule of dimension exactly equal to $N$.However the
definition of this last submodule (up to equivalence) depends on the choice
of an arbitrary complex parameter $\lambda$, so that we have a chain of
inclusions for every such parameter.

%%%%%%%%%%%%%%%%%%%%%%%%%%%%%%%%%%%%%%%%%%%%%%%%%%%%%%%%%%%%%%%%%%%%%%%%%%

\subsection{Decomposition of ${\mathcal M} = M_N(\CC)$ in representations
            of $\mathcal H$}
\label{subsec:M-in-reps-H}

Since there is an action of $\mathcal H$ on $\mathcal M$, it is clear
that $\mathcal M$, as a vector space, can be analysed in terms of
representations of $\mathcal H$. The following result was shown in
\cite{Coquereaux-Schieber}:

Under the left action of $\mathcal{H}$, the algebra of $N\times N$
matrices can be decomposed into a direct sum of invariant subspaces of
dimension $N$, according to
\begin{itemize}
\item $N_N = N_{irr}$: irreducible
\item $N_{N-1}$: reducible indecomposable, with an invariant subspace
      of dimension $N-1$.
\item $N_{N-2}$: reducible indecomposable, with an invariant subspace
      of dimension $N-2$.
\item $\vdots$
\item $N_{1}$: reducible indecomposable, with an invariant subspace
      of dimension $1$.
\end{itemize}
The elements of the module called $N_{p}$ (of dimension $N$) are of the
kind:
$$
{\underline{N}}_{p} = \begin{pmatrix}{
      \gamma_1\theta_{\lambda_1} & \gamma_2\theta_{\lambda_2} & \cdots &
      \gamma_{N-p}\theta_{\lambda_{N-p}} & \beta_1\theta_1\theta_2 &
      \beta_2\theta_1\theta_2 & \cdots & \beta_{p}\theta_1\theta_2}
   \end{pmatrix}
$$

\noindent This submodule is the direct sum of an invariant sub-module of
dimension $p$, and a vector subspace of dimension $N-p$
$$
   {\underline{N}}_{p} = \underline{p} \inplus (N-p)
$$
with
$$
\underline{p} = \begin{pmatrix}{ 0 & 0 & \cdots & 0 &
       \beta_1\theta_1\theta_2 & \beta_2\theta_1\theta_2 & \cdots &
       \beta_{p}\theta_1\theta_2}
    \end{pmatrix}
$$

Using these notations, the algebra of complex matrices $N \times N$ can be
written
$$
   \mathcal{M} = N_N \oplus N_1 \oplus N_2 \oplus \cdots \oplus N_{N-1}
$$
In the particular case $N=3$, $3_{odd}$ is an abelian subalgebra of
$\mathcal M$, actually isomorphic with the algebra $\CC[\ZZ_3]$ of the
abelian group $\ZZ_3$. Hence, we may write
$$
   {\mathcal M} = \CC[\ZZ_3] \oplus x \, \CC[\ZZ_3] \oplus x^2 \, \CC[\ZZ_3]
                = 3_{odd} \oplus 3_{eve} \oplus 3_{irr} \ .
$$
Moreover, it can be shown that
$$
   Inverse(3_{eve}) \subset 3_{irr} \ , \quad
   Inverse(3_{irr}) \subset 3_{eve} \ , \quad \text{but} \quad
   Inverse(3_{odd}) \subset 3_{odd} \ .
$$

%%%%%%%%%%%%%%%%%%%%%%%%%%%%%%%%%%%%%%%%%%%%%%%%%%%%%%%%%%%%%%%%%%%%%%%%%%

\subsection{The universal $R$-matrix}

The finite dimensional Hopf algebra $\mathcal H$ we have been studying is
actually braided and quasi-triangular (as it is well known, the quantum
enveloping algebra of $SL(2)$ does {\sl not} possess these properties
when $q$ is specialized to a root of unity). The $R$-matrix of $\mathcal H$
can be obtained directly from a general formula given by \cite{Rosso} but we
can also get it in a pedestrian way by starting from a reasonable ansatz and
imposing certain conditions. Here we take $N=3$. We start from
$$
   R = R_K R_X
$$
where
\begin{eqnarray*}
   R_K &=& \sum_{i,j = 0,1,2} c_{ij} K^i \otimes K^j \\
   R_X &=& \left[ \one \otimes \one + \alpha X_- \otimes X_+
                  + \beta X_-^2 \otimes X_+^2 \right]
\end{eqnarray*}
Here $\alpha$, $\beta$ and the $c_{ij}$ are complex numbers (symmetric in
$i,j$).

A quasi-triangular R matrix should satisfy $(\epsilon \otimes \id)R = \one$.
As $\epsilon(K)=1$, this condition implies that
\begin{eqnarray*}
R_K &=& (1-c_{01}-c_{02}) \one \otimes \one
         + c_{01} (\one \otimes K + K \otimes \one)
         + c_{02} (\one \otimes K^2 + K^2 \otimes \one) \\
    & &  + c_{12} (K \otimes K^2 + K^2 \otimes K)
         - (c_{01}+c_{12}) K \otimes K
         - (c_{02}+c_{12}) K^2 \otimes K^2 \ .
\end{eqnarray*}
Also, we should have $(S \otimes S)R = R$. Comparing terms with zero
$X_\pm$ grading (\ie with no $X_\pm$ terms) we find
$(S \otimes S)R_K = R_K$, and thus $c_{01} = c_{02}$. Making use of the
terms in $X_- \otimes X_+$ we get $c_{01} = 1/3$, $c_{12} = q^2/3$. In
the same way, one finds $\alpha = (q-q^{2})$ and $\beta = 3q$.

The universal $R$-matrix now reads explicitly
\begin{eqnarray}
R &=& \frac{1}{3} \left[
           \one \otimes \one + (\one \otimes K + K \otimes \one)
           + (\one \otimes K^2 + K^2 \otimes \one) \right. \nonumber \\
  & & \quad \left. + q^2 (K \otimes K^2 + K^2 \otimes K)
           + q K \otimes K + q K^2 \otimes K^2 \right]     \\
  & & \times \left[ \one \otimes \one + (q-q^{-1}) X_- \otimes X_+
           + 3q X_-^2 \otimes X_+^2 \right]                \nonumber
\label{R-matrix}
\end{eqnarray}
Using the explicit numerical matrices $X_\pm, K$ given in Appendix~E of
\cite{CoGaTr-E}, one can obtain the numerical $R$ matrices in various
representations of interest (irreducible or indecomposable ones).

Note that $R^{-1} = R_X^{-1} R_K^{-1}$, where
\begin{eqnarray*}
   R_X^{-1} = \left[ \one \otimes \one - \alpha X_- \otimes X_+
                   + (\alpha^2 - \beta) X_-^2 \otimes X_+^2 \right]
\end{eqnarray*}
and $R_K^{-1}$ is given by the same expression as $R_K$ but with $q$ and
$q^2$ interchanged. Here we already see that our algebra is {\em not}
triangular, $R_{21}$ (the flipped $R$) has terms of the form
$X_+ \otimes X_-$, whereas $R^{-1}$ only contains terms of the form
$X_- \otimes X_+$. It can be straightforwardly verified (but it is
cumbersome) that the requirements of almost-cocommutativity and
quasi-triangularity hold, namely
\begin{eqnarray}
   \Delta^{op}(h) = R \Delta(h) R^{-1} \qquad h \in \mathcal H \ ,
\end{eqnarray}
and
\begin{eqnarray}
   (\Delta \otimes \id)(R) &=& R_{13} R_{23} \nonumber \\
   (\id \otimes \Delta)(R) &=& R_{13} R_{12}
\end{eqnarray}

\noindent For a generic (odd) $N$ the universal $R$-matrix reads
\begin{eqnarray*}
R &=& \frac{1}{N} \left[
        \sum_{0 \le m,n < N} q^{mn} \: K^m \otimes K^n \right]
      \left[ \sum_{0 \le k < N} {1\over [k]_q !}
        (1-q^{-2})^k q^{k(k+1)/2} \: X_-^k \otimes X_+^k \right]
\end{eqnarray*}

In the case $N=3$, the algebra $\mathcal H$ has three projective
indecomposable modules of dimensions denoted $3_{irr}$, $6_{odd}$ and
$6_{eve}$. The first one is irreducible whereas the last two are not.
The quotient of $6_{odd}$ (respectively $6_{eve}$) by their radical
of respective dimensions $5$ and $4$ give irreducible representations
of dimensions $1$ and $2$. Moreover, the tensor products between
irreducible representations and projective indecomposable ones can be
reduced as follows:
$$
\begin{tabular}{llllll}
$\underline{2} \times \underline{2}$ & $\equiv$ &
   $\underline{1} + \underline{3}_{irr}$ &
$\underline{6}_{eve} \times \underline{3}_{irr}$ & $\equiv$ &
   $\underline{6}_{eve} + \underline{6}_{eve} + \underline{3}_{irr} +
      \underline{3}_{irr}$ \\
$\underline{2} \times \underline{3}_{irr}$ & $\equiv$ &
   $\underline{6}_{eve}$ &
$\underline6_{odd} \times \underline{3}_{irr}$ & $\equiv$ &
   $\underline{6}_{eve} + \underline{6}_{eve} + \underline{3}_{irr} +
      \underline{3}_{irr}$ \\
$\underline{3}_{irr} \times \underline{3}_{irr}$ & $\equiv$ &
   $\underline{6}_{odd} + \underline{3}_{irr}$ &
$\underline{6}_{eve} \times \underline{6}_{eve}$ & $\equiv$ &
   $4(\underline{6}_{eve}) + 4(\underline{3}_{irr})$ \\
$\underline{6}_{eve} \times \underline{2}$ & $\equiv$ &
   $\underline{6}_{odd} + \underline{3}_{irr} + \underline{3}_{irr}$ &
$\underline{6}_{eve} \times \underline{6}_{odd}$ & $\equiv$ &
   $4(\underline{6}_{eve}) + 4(\underline{3}_{irr})$ \\
$\underline{6}_{odd} \times \underline{2}$ & $\equiv$ &
   $\underline{6}_{eve} + \underline{3}_{irr} + \underline{3}_{irr}$ &
$\underline{6}_{odd} \times \underline{6}_{odd}$ & $\equiv$ &
   $2(\underline{6}_{odd}) + 2(\underline{6}_{eve}) +
      4(\underline{3}_{irr})$
\end{tabular}
$$
Notice that products of irreducible representations are not always direct
sums of irreducibles (this is not a modular category). One can define a
concept of truncated (or fusion) tensor product by using the notion of
quantum trace and discarding those representations of $q$-dimension zero.
The algebra $\mathcal H$ is indeed a Ribbon Hopf algebra and the notion of
quantum trace (and of quantum dimension of representations) makes sense.
This quantum dimension $Tr_q$ has the property of being multiplicative with
respect to tensor products. It can be seen that $Tr_q(X) = Tr(KX)$ so that
the $q$-dimension of the projective indecomposable representations vanishes,
whereas it is equal to the $q$-number $[n]$ for the irreducible
representations of (usual) dimensions $n$. Notice that for each value of
$q$ being a primitive $N$-th root of unity ($N$ odd), there exists a
particular projective indecomposable representation of (usual) dimension
$N$ which is, at the same time, irreducible; the $q$ dimension of this
particular irreducible representation vanishes. For $N = 3$, for instance,
one can check, using the Appendix~E of \cite{CoGaTr-E} that the
$q$-dimension of the projective indecomposable representations (the
$3_{irr}$, $6_{eve}$ and $6_{odd}$) vanishes, whereas it is equal
respectively to $1$ and $-1$ for the irreducible representations of
(usual) dimensions $1$ and $2$.

The previous table of results for tensor products of representations of
$\mathcal H$ was obtained in \cite{CoGaTr-E} without using knowledge of
the $R$-matrix and without using the concept of $q$-dimension (or truncated
tensor product).

%%%%%%%%%%%%%%%%%%%%%%%%%%%%%%%%%%%%%%%%%%%%%%%%%%%%%%%%%%%%%%%%%%%%%%%%%%

\section{Reality structures}
\label{sec:stars}

%%%%%%%%%%%%%%%%%%%%%%%%%%%%%%%%%%%%%%%%%%%%%%%%%%%%%%%%%%%%%%%%%%%%%%%%%%

\subsection{Real forms and stars on quantum groups}

A $*$-Hopf algebra $\mathcal F$ is an associative algebra that satisfies the
following properties (for all elements $a, b$ in $\mathcal F$):

\begin{enumerate}
\item
   $\mathcal F$ is a Hopf algebra (a quantum group), with coproduct
   $\Delta$, antipode $S$ and counit $\epsilon$.

\item
   $\mathcal F$ is an involutive algebra, \ie it has an involution $*$
   (a `star' operation). This operation is involutive ($**a = a$),
   antilinear ($*(\lambda a) = \overline{\lambda} *a$, where $\lambda$ is
   a complex number), and anti-multiplicative ($*(ab) = (*b)(*a)$).

\item
   The involution is compatible with the coproduct, in the following sense:
   if $\Delta a = a_1 \otimes a_2$, then $\Delta *a = *a_1 \otimes *a_2$.

\item
   The involution is also compatible with the counit:
   $\epsilon(*a) = \overline{\epsilon(a)}$.

\item
   The following relation with the antipode holds: $S * S * a = a$.
\end{enumerate}

\noindent Actually, the last relation is a consequence of the others. It
can also be written $ S \, * = * \, S^{-1}$. It may happen that $S^2 = \id$,
in which case $S \, * = * \, S$, but it is not generally so (usually the
square of the antipode is only a conjugacy).

If one wishes, using the $*$ on $\mathcal F$, one can define a star
operation on the tensor product ${\mathcal F} \otimes {\mathcal F}$, by
$*(a \otimes b) = *a \otimes *b$. The third condition reads then
$\Delta \, * = * \, \Delta$, so one can say $\Delta$ is a $*$-homomorphism
between $\mathcal F$ and ${\mathcal F} \otimes {\mathcal F}$ (each with its
respective star). It can also be said that $\epsilon$ is a $*$-homomorphism
between $\mathcal F$ and $\CC$ with the star given by complex conjugation.

A star operation as above, making the Hopf-algebra a $*$-Hopf algebra,
is also called a {\sl real form\/} for the given algebra. An element
$u$ that is such that $* u = u$ is called a real element.

%%%%%%%%%%%%%%%%%%%%%%%%%%%%%%%%%%%%%%%%%%%%%%%%%%%%%%%%%%%%%%%%%%%%%%%%%%

\subsubsection{Twisted stars on quantum groups}

It may happen that one finds an involution $*$ on a Hopf algebra for
which the third axiom fails in a special way, namely, the case where
$\Delta a = a_1 \otimes a_2$ but where $\Delta *a = *a_2 \otimes *a_1$.
In this case $S \, * = * \, S$. Such an involution is called a
{\sl twisted star\/} operation. Remember that, whenever one has a coproduct
$\Delta$ on an algebra, it is possible to construct another coproduct
$\Delta^{op}$ by composing the first one with the tensorial flip. If
one defines a star operation on the tensor product (as before) by
$*(a \otimes b) \doteq *a \otimes *b$, the property defining a twisted
star reads
$$
   \Delta \, * = * \, \Delta^{op} \ .
$$

One should be cautious: some authors introduce a different star operation
on the tensor product, namely $*'(a \otimes b) \doteq *b \otimes *a$. In
that case, a twisted star operation obeys $\Delta \, * = *' \, \Delta$!
Twisted star operations are often used in conformal field theory
(\cite{Mack}).

%%%%%%%%%%%%%%%%%%%%%%%%%%%%%%%%%%%%%%%%%%%%%%%%%%%%%%%%%%%%%%%%%%%%%%%%%%
\subsubsection{Remark about superstars on differential algebras}

On a real manifold, the star operation has a ``strange property''.
Indeed, it is natural to take $*x = x$, $*y = y$ (on the algebra of
functions) and extend it to the algebra of differential forms by requiring
that the star is that $*dx = dx$, so that it is ``real'' in the sense of
not being complex. However, antimultiplicativity of the star leads
immediately to $*(dx \, dy) = (*dy) \, (*dx) = dy \, dx = - dx \, dy$, so
that $dx \, dy$ cannot be a ``real element'' for this reasonable star!
This strange feature does not arise when we stop at the level of the
algebra of functions but it shows up as soon as we want to promote a given
star to the $\ZZ_{2}$-graded algebra of differential forms, something that
we shall do later in our context. In order to solve this ``problem'',
which already appears on a usual manifold, it is always possible ---but
not necessary--- to introduce a superstar (nothing to do with the twist
described in the previous subsection), \ie a $\ZZ_{2}$-graded star, with
the constraint:
$$
   *(a \, b) = (-1)^{(\#a \#b)} \, *b \, *a
$$
where $\#a$ is the $\ZZ_{2}$-parity of $a$. Using such superstars allow one
to identify ``real elements'' (in the usual sense of not being complex) with
real elements for the $*$ (\ie such that $*u=u$).

%%%%%%%%%%%%%%%%%%%%%%%%%%%%%%%%%%%%%%%%%%%%%%%%%%%%%%%%%%%%%%%%%%%%%%%%%%

\subsection{Real forms on $\mathcal F$}

As can be easily found in the literature, one has three possibilities
for the ---not twisted--- star operations on $Fun(SL_q(2,\CC))$ (up to
Hopf automorphisms):

\begin{enumerate}
\item
   The real form $Fun(SU_q(2))$. The matrix elements obey
   $a^* = d$, $b^* = -qc$, $c^* = -q^{-1} b$ and $d^* = a$. Moreover,
   $q$ should be real.

\item
   The real form $Fun(SU_q(1,1))$. The matrix elements obey
   $a^* = d$, $b^* = qc$, $c^* = q^{-1} b$ and $d^* = a$. Moreover,
   $q$ should be real.

\item
   The real form $Fun(SL_q(2,\RR))$. The matrix elements obey
   $a^* = a$, $b^* = b$, $c^* = b$ and $d^* = d$.
   Here $q$ can be complex but it should be a phase.
\end{enumerate}

\noindent Therefore, taking $q$ a root of unity is incompatible with the
$SU_q$ real forms, and the only possibility is to choose the star
corresponding to $Fun(SL_q(2,\RR))$. This already tells us that there
is at most one real form on its quotient $\mathcal F$. We only have to
check that the star operation preserves the ideal and coideal defined by
$a^3 = d^3 = \one$, $b^3 = c^3 = 0$. This is trivial because $a^* = a$,
$b^* = b$, $c^* = c$ and $d^* = d$. Hence, this star operation can be
restricted to $\mathcal F$.

This real form can be considered as a reduced quantum group associated
with the real form $Fun(SL_q(2,\RR))$ of $Fun(SL_q(2,\CC))$.

Of course, one can also discuss {\sl twisted} star operations: see,
in particular the comment at the end of the next subsection.

%%%%%%%%%%%%%%%%%%%%%%%%%%%%%%%%%%%%%%%%%%%%%%%%%%%%%%%%%%%%%%%%%%%%%%%%%%

\subsection{Real structures and star operations on $\mathcal M$
            and $\mathcal H$}

Now we want to introduce an involution (a star operation) on the
comodule algebra $\mathcal M$. This involution should be compatible with
the coaction of $\mathcal F$. That is, we are asking for covariance of
the star under the (right,left) $\mathcal F$-coaction,
\begin{equation}
   (\Delta_{R,L} \, z)^* = \Delta_{R,L} (z^*) \ , \quad
                           \text{for any} z \in {\mathcal M} \ .
\label{*-delta-condition}
\end{equation}

\noindent Here we have used the same notation $*$ for the star on the
tensor products, which are defined as $(A \otimes B)^* = A^* \otimes B^*$.
Using, for instance, the left coaction in (\ref{*-delta-condition}),
we see immediately that the real form $Fun(SL_q(2,\RR))$ corresponds
to choosing on ${\mathcal M} = M_N(\CC)$ the star given by
\begin{equation}
   x^* = x \ , \qquad y^* = y \ .
\label{*-on-M}
\end{equation}

We now want to find a compatible $*$ on the algebra $\mathcal H$.
As $\mathcal H$ is dual to $\mathcal F$ (or $U_q(sl(2))$ dual to
$Fun(SL_q(2,\CC))$), we should have dual $*$-structures. This means
the relation
\begin{equation}
   <h^*, u> = \overline{<h, (Su)^*>} \ , \quad
                 h \in{\mathcal H}, u \in{\mathcal F}
\label{dual-*-structures}
\end{equation}
holds. In this way we obtain:
\begin{equation}
   X_+^* = -q^{-1} X_+ \ , \quad X_-^* = -q X_- \ , \quad K^* = K \ .
\label{*-on-H}
\end{equation}

Moreover, the covariance condition for the star, equation
(\ref{*-delta-condition}), may also be written dually as a condition for
$*$ under the action of $\mathcal H$. This can be done pairing the
$\mathcal F$ component of equation (\ref{*-delta-condition}) with some
$h \in {\mathcal H}$. One gets finally the constraint on
$*_{\mathcal H}$ to be $\mathcal H$ covariant,
\begin{equation}
   h(z^*) = \left[ (S h )^* z \right]^* \ , \quad
                  h \in{\mathcal H}, z \in{\mathcal M} \ .
\label{*-action-condition}
\end{equation}

Adding the non quadratic relations $x^N = \one$, $y^N = \one$ in
$\mathcal M$, and the corresponding ones in the algebra $\mathcal H$,
does not change anything to the determination of the star structures.
This is because the (co)ideals are preserved by the involutions, and
thus the quotients can still be done.

Remark that the set of $N \times N$ matrices is endowed with a usual
star operation, the hermitian conjugacy $\dag$. It is clear that $x$
and $y$ are unitary elements with respect to $\dag$: $x^\dag = x^{-1}$
($=x^{2}$ if $N=3$) and $y^\dag = y^{-1}$ ($=y^{2}$ if $N=3$). But
this is {\sl not} the star operation that we are using now, the one that
is compatible with the quantum group action, at least when $x$ and $y$
are represented by the $3 \times 3$ matrices given in
Section~\ref{sec:red-q-plane}.

Note finally that one could be tempted to chose the involution defined
by $K^\star = K^{-1}$, $X_+^\star = \pm X_-$ and $X_-^\star = \pm X_+$.
However, this is a {\em twisted} star operation. This last operation is
the one one would need to interpret the unitary group of $\mathcal H$,
in the case $N=3$, as $U(3) \times U(2) \times U(1)$, which could be
related to the gauge group of the Standard Model \cite{Connes-2}.

%%%%%%%%%%%%%%%%%%%%%%%%%%%%%%%%%%%%%%%%%%%%%%%%%%%%%%%%%%%%%%%%%%%%%%%%%%

\subsection {A quantum group invariant scalar product on the space of
             $N \times N$ matrices}
\label{subsec:scalar-product-on-M}

\subsubsection{Identification between star operation and adjoint}

A possible scalar product on the space ${\mathcal M} = M_N(\CC)$ of
$N \times N$ matrices is the usual one, namely, $m_1, m_2 \rightarrow
Tr(m_1^\dag m_2)$. For every linear operator $\ell$ acting on $\mathcal M$
we can define the usual adjoint $\ell^\dag$; however, this adjoint does not
coincide with the star operation introduced previously. Our aim in this
section is to find another scalar product better suited for our purpose.

We take $z, z' \in {\mathcal M}$, and $h \in \mathcal H$. We know that the
first ones act on $\mathcal M$ like multiplication operators and that $h$
acts on $\mathcal M$ by twisted derivations or automorphisms. We also
know the action of our star operation $*$ on these linear operators. We
shall now obtain our scalar product by imposing that $*$ coincides with the
adjoint associated with this scalar product (compatibility condition). That
is, we are asking for an inner product on $\mathcal M$ such that the actions
of $\mathcal M$ and $\mathcal H$ (each with its respective star) on that
vector space may be thought as $*$-representations. Due to the fact that
$(z, z') = (\one, z^* z')$, it is enough to compute $(\one, z)$ for all the
$z$ belonging to $\mathcal M$. The above compatibility condition leads to a
single solution \cite{CoGaTr-E}: the only non vanishing scalar product
between elements of the type $(\one, z)$ is $(\one, x^{N-1} y^{N-1})$.
From this quantity we deduce $N^{2}-1$ other non-zero values for the scalar
products $(z,z')$ where $z$ and $z'$ are basis elements $x^r y^s$. For
instance,
$(x, x^{N-2}y^2) = (\one, x^* x^{N-2} y^{N-1}) = (\one, x^{N-1} y^{N-1})$.

Hermiticity with respect to $*$ implies that $(xy,xy)$ should be real,
so we set $(xy,xy) = 1$.

%%%%%%%%%%%%%%%%%%%%%%%%%%%%%%%%%%%%%%%%%%%%%%%%%%%%%%%%%%%%%%%%%%%%%%%%%%

\subsubsection{Quantum group invariance of the scalar product}

We should now justify why the above scalar product was called a quantum
group invariant one. Remember we only said the scalar product was such that
the stars coincide with the adjoint operators, or such that the actions are
given by $*$-representations.

We refer the reader to \cite{CoGaTr}, where it is shown that the
$*$-representation condition on the scalar product,
\begin{equation}
   (h z,w) = (z,h^* w) \ , \quad h \in {\mathcal H} \ ,
\label{star_rep}
\end{equation}
automatically fulfills one of the two alternative invariance conditions
that can be imposed on the scalar product, namely
\begin{equation}
   ((Sh_1)^* z,h_2 w) = \epsilon(h) (z,w) \ , \quad
                        \text{with} \Delta h = h_1 \otimes h_2 \ .
\label{cond_for_scalar_prod_inv}
\end{equation}

The relations dual to (\ref{cond_for_scalar_prod_inv}) and
(\ref{star_rep}) are those that apply to the coaction of $\mathcal F$
instead of the action of $\mathcal H$. These are
\begin{equation}
   (\Delta_R \, z, \Delta_R \, w) = (z,w) \one_{\mathcal F} \ ,
\label{dual_cond_for_scalar_prod_inv}
\end{equation}
where $(\Delta_R \, z, \Delta_R \, w)$ should be understood as
$(z_i, w_j) \, T_i^* \, T_j$ if $\Delta_R \, z = z_i \otimes T_i$, and
\begin{equation}
   (z,\Delta_R w) = ((1\otimes S)\Delta_R z,w) \ ,
\label{dual_star_rep}
\end{equation}
respectively. We have used here the right-coaction, but the formulas for
the left coaction can be trivially deduced from the above ones.

It is worth noting that these equations for the coaction of $\mathcal F$
{\em imply} the previous ones for the action of $\mathcal H$, and are
completely equivalent assuming non-degeneracy of the pairing
$<\cdot,\cdot>$. Moreover, (\ref{dual_cond_for_scalar_prod_inv}) is a
requirement analogous to the condition of classical invariance by a group
element action.

Using the unique Hopf compatible star operation $*$ on $\mathcal H$, we
can calculate the most general metric on the vector spaces of each of the
indecomposable representations of $\mathcal H$. Obviously, one should
restrict the inner product to be a quantum group invariant one. This is done
in \cite{CoGaTr-E}, where we refer for details. Here it suffices to say that
one obtains for the projective indecomposable modules nondegenerate
but indefinite metrics, and the submodules carry a metric which is,
moreover, degenerate.

%%%%%%%%%%%%%%%%%%%%%%%%%%%%%%%%%%%%%%%%%%%%%%%%%%%%%%%%%%%%%%%%%%%%%%%%%%

\section{The Manin dual ${\mathcal M}^!$ of $\mathcal M$}
\label{sec:manin-dual}

Our algebra $\mathcal M$ is not quadratic since we impose the
relations $x^N = y^N = \one$. Nevertheless, we define its Manin dual
${\mathcal M}^!$ as the algebra generated {\sl over the complex
numbers\/} by $\xi^1, \xi^2$, satisfying the relations
$$
   (\xi^1)^2 = 0                  \: , \quad
   q\xi^1 \xi^2 + \xi^2 \xi^1 = 0 \: , \quad
   (\xi^2)^2 = 0                  \ ,
$$
as in the unreduced case. We shall write $dx \doteq \xi^1$ and
$dy \doteq \xi^2$, so ${\mathcal M}^!$ is defined by these two
generators and the relations
\begin{equation}
   dx^2 = 0                  \: , \quad
   dy^2 = 0                  \: , \quad
   q dx \, dy + dy \, dx = 0 \ .
\label{M!-relations}
\end{equation}

Once the coaction of $\mathcal F$ on $\mathcal M$ has been defined as
in Section~\ref{sec:q-group-F}, it is easy to check that its coaction
on ${\mathcal M}^!$ is given by the same formulae. Namely, writing
$$
   \pmatrix{dx' \cr dy'} =
      \pmatrix{a & b \cr c & d} \otimes \pmatrix{dx \cr dy}
$$
and
$$
   \pmatrix{\tilde dx & \tilde dy} =
      \pmatrix{dx & dy} \otimes \pmatrix{a & b \cr c & d}
$$
ensures that $q \, dx' \, dy' + dy' \, dx' = 0$ and
$q \, \tilde dx \, \tilde dy + \tilde dy \, \tilde dx = 0$,
once the relation $q \, dx \, dy + dy \, dx = 0$ is satisfied.
The left and right coactions can be read from those formulae, for
instance $\Delta_R (dx) = dx \otimes a + dy \otimes c$.

Since the formulae for the coactions on the generators and on their
differentials are the same, the formulae for the actions of $\mathcal H$
on ${\mathcal M}^!$ must also coincide. For instance, using $X_-^L[x] = y$
we find immediately $X_-^L[dx] = dy$. This corresponds to an irreducible
two dimensional representation of $\mathcal H$. We shall return to this
problem in the next section, since we are going to analyse the
decomposition in representations of a differential algebra
$\Omega_{WZ}({\mathcal M})$ built using $\mathcal M^!$.

%%%%%%%%%%%%%%%%%%%%%%%%%%%%%%%%%%%%%%%%%%%%%%%%%%%%%%%%%%%%%%%%%%%%%%%%%%
%%%%%%%%%%%%%%%%%%%%%%%%%%%%%%%%%%%%%%%%%%%%%%%%%%%%%%%%%%%%%%%%%%%%%%%%%%

\section{Covariant differential calculus on $\mathcal M$}
\label{sec:diff-calculus}

Given an algebra $\mathcal A$, there is a universal construction that
allows one to build the so-called {\sl algebra of universal differential
forms\/} $\Omega({\mathcal A}) = \sum_{p=0}^\infty \Omega^p({\mathcal A})$
over $\mathcal A$. This differential algebra is universal, in the sense that
any other differential algebra with $\Omega^0({\mathcal A}) = {\mathcal A}$
will be a quotient of the universal one. For practical purposes, it is often
not very convenient to work with the algebra of universal forms. First of
all, it is very ``big''. Next, it does not remember anything of the coaction
of ${\mathcal F}$ on the algebra ${\mathcal M}$ (the $0$-forms).

Starting from a given algebra, there are several constructions that allow
one to build ``smaller'' differential calculi. As already mentioned, they
will all be quotients of the algebra of universal forms by some
(differential) ideal. One possibility for such a construction was
described by \cite{Connes}, another one by \cite{Dubois-Violette},
and yet another one by \cite{Coquereaux-Haussling-Scheck}. In the present
case, however, we use something else, namely the differential calculus
$\Omega_{WZ}$ introduced by Wess and Zumino \cite{Wess-Zumino} in the case
of the quantum $2$-plane. Its main interest is that it is covariant with
respect to the action (or coaction) of a quantum group. Its construction
was recalled in \cite{CoGaTr} where it was also shown that one can
further take another quotient by a differential ideal associated with the
constraints $x^{N}=y^{N}=\one$ (so that, indeed $d(x^{N})=d(y^{N}) = 0$
automatically).

%%%%%%%%%%%%%%%%%%%%%%%%%%%%%%%%%%%%%%%%%%%%%%%%%%%%%%%%%%%%%%%%%%%%%%%%%%

\subsection{The reduced Wess-Zumino complex}

The algebra $\Omega_{WZ}$ is a differential algebra first
introduced by \cite{Wess-Zumino} for the ``full'' quantum plane.
First of all
$\Omega_{WZ} = \Omega_{WZ}^0 \oplus \Omega_{WZ}^1 \oplus \Omega_{WZ}^2$
is a graded vector space.

\begin{itemize}
\item
   Forms of grade $0$ are just functions on the quantum plane,
   \ie elements of $\mathcal M$.

\item
   Forms of grade $1$ are of the type
   $a_{rs} x^r y^s dx + b_{rs} x^r y^s dy$, where $dx$ and $dy$ are the
   generators of the Manin dual ${\mathcal M}^!$.

\item

Forms of grade $2$ are of the type $c_{rs} x^r y^s dx \, dy$.
\end{itemize}

\noindent Next, $\Omega_{WZ}$ is an algebra. The relations between $x$, $y$,
$dx$ and $dy$ are determined by covariance under the quantum group action:

\begin{equation}
\begin{tabular}{ll}
   $xy = qyx$                 &                                        \\
   $x\,dx = q^2 dx\,x$ \qquad & $x\,dy = q \, dy\,x + (q^2 - 1) dx\,y$ \\
   $y\,dx = q \, dx\,y$       & $y\,dy = q^2 dy\,y$                    \\
   $dx^2 = 0$                 & $dy^2 = 0$                             \\
   $dx\,dy + q^2 dy\,dx = 0$  &                                        \\
\end{tabular}
\label{WZ-relations}
\end{equation}

Moreover, we want $\Omega_{WZ}$ to be a differential algebra, so we
introduce an operator $d$ and set $d(x) = dx$, $d(y) = dy$; the Leibniz rule
(for $d$) should also hold. Finally, we impose $d \one = 0$ and $d^2 = 0$.

In the case $q^N = 1$, we add to $\Omega_{WZ}$ the extra defining relation
(coming from the definition of the reduced quantum plane):
$x^N = y^N = \one$. The fact that $\Omega_{WZ}$ is still well defined as a
differential algebra is not totally obvious and requires some checking (see
\cite{CoGaTr-E}). Note that $\dim(\Omega_{WZ}^0) = N^{2}$,
$\dim(\Omega_{WZ}^1) = 2N^{2}$ and $\dim(\Omega_{WZ}^2) = N^{2}$.

%%%%%%%%%%%%%%%%%%%%%%%%%%%%%%%%%%%%%%%%%%%%%%%%%%%%%%%%%%%%%%%%%%%%%%%%%%

\subsection{The action of $\mathcal H$ on $\Omega_{WZ}({\mathcal M})$}
\label{subsec:H-action-on-Omega}

Since $\mathcal H$ acts on ${\mathcal M}$ (and we know how this module
decomposes under the action of $\mathcal H$), it is clear that we can also
decompose $\Omega_{WZ}$ in representations of $\mathcal H$. This was done
explicitly (for the case $N=3$) in \cite{CoGaTr}, and the action of
$\mathcal H$ on $\Omega_{WZ}^1$ (for an arbitrary $N$) was described in
\cite{Coquereaux-Schieber}. The cohomology of $d$ is actually non trivial
and was also studied in \cite{CoGaTr-E}.

%%%%%%%%%%%%%%%%%%%%%%%%%%%%%%%%%%%%%%%%%%%%%%%%%%%%%%%%%%%%%%%%%%%%%%%%%%

\subsection{The space of differential operators on $\mathcal M$}
\label{app:diff-operators-on-M}

We now summarize the structure of the space of differential operators on
the reduced quantum plane $\mathcal M$, \ie the algebra of $N \times N$
matrices.

%%%%%%%%%%%%%%%%%%%%%%%%%%%%%%%%%%%%%%%%%%%%%%%%%%%%%%%%%%%%%%%%%%%%%%%%%%

\subsubsection*{The space $\mathcal D$ of differential operators on
                $\mathcal M$}

We already know what the operator
$d: \Omega_{WZ}^0 = {\mathcal M} \rightarrow \Omega_{WZ}^1$ is.
{\it A priori\/} $d f = dx \, \partial_x( f) + dy  \, \partial_y (f)\:$,
where $\partial_x f, \partial_y f \in {\mathcal M}$, and this defines
$\partial_x$ and $\partial_y$ as (linear) operators on ${\mathcal M}$.

Generally speaking, operators of the type $f(x,y) \partial_x$ or
$f(x,y) \partial_y$ are called differential operators of order $1$.
Composition of such operators gives rise to differential operators of
order higher than $1$. Multiplication by elements of $\mathcal M$ is
considered as a differential operator of degree $0$. The space of all
these operators is a vector space
${\mathcal D} = \oplus_{i=0}^{i=4}{\mathcal D}_i$.

%%%%%%%%%%%%%%%%%%%%%%%%%%%%%%%%%%%%%%%%%%%%%%%%%%%%%%%%%%%%%%%%%%%%%%%%%%

\subsubsection*{The twisting automorphisms $\sigma$\footnote{
                   Some properties of these automorphisms are discussed
                   in \cite{Manin-2}
               }}

Since we know how to commute $x,y$ with $dx,dy$, we can write, for any
element $f \in {\mathcal M}$
\begin{eqnarray*}
   f dx &=& dx \, \sigma_x^x(f) + dy \, \sigma_y^x(f) \cr
   f dy &=& dx \, \sigma_x^y(f) + dy \, \sigma_y^y(f)
\end{eqnarray*}
where each $\sigma_{j}^{i}$ is an element of $End({\mathcal M})$ to be
determined (just take $f=x$ and $f=y$ in the above to get the results that
one is looking for).

Let $f$ and $g$ be elements of $\mathcal M$. From the associativity
property $(fg) dz = f(g dz)$ we find
$$
   \sigma_i^j(fg) = \sigma_i^k(f) \, \sigma_k^j(g)
$$
with a summation over the repeated indices. The map
$\sigma: f \in {\mathcal M} \rightarrow \sigma(f) \in M_2({\mathcal M})$
is an algebra homomorphism from the algebra ${\mathcal M}$ to the algebra
$M_2({\mathcal M})$ of $2 \times 2$ matrices, with elements in $\mathcal M$.

The usual Leibniz rule for $d$, namely $d(fg) = d(f)g+fd(g)$, implies
that
$$
 \partial_i(fg) = \partial_i(f) \, g + \sigma_i^j (f) \, \partial_j(g) \ .
$$
This shows that $\partial_x$ and $\partial_y$ are derivations twisted
by an automorphism.

%%%%%%%%%%%%%%%%%%%%%%%%%%%%%%%%%%%%%%%%%%%%%%%%%%%%%%%%%%%%%%%%%%%%%%%%%%

\subsubsection*{Relations in $\mathcal D$}

For calculational purposes, it is useful to know the commutation relations
between $x,y$ and $\partial_x, \partial_y$, those between
$\partial_x, \partial_y$ and $\sigma_i^j$ and the relations between the
$\sigma_i^j$. Here are some results (see also \cite{Wess-Zumino,Manin}).
\begin{eqnarray*}
   \partial_x\, x &=& 1 + q^2 x\, \partial_x + (q^2-1) y \, \partial_y \cr
   \partial_x\, y &=& q y \, \partial_x \cr
               {} &{}& \cr
   \partial_y\, x &=& q x \, \partial_y \cr
   \partial_y\, y &=& 1 + q^2 y \, \partial_y
\end{eqnarray*}
Moreover,
$$
   \partial_y \, \partial_x = q \partial_x \, \partial_y
$$

Also, the relations $x^N = y^N = \one$ lead to other constraints on the
powers of the derivations. For example, for $N=3$ these imply the
constraint:
$$
   \partial_x\partial_x\partial_x = \partial_y\partial_y\partial_y = 0
$$
Finally, the commutation relations between the $\sigma$'s can be calculated
from the values of the $\sigma^i_j(x)$.

%%%%%%%%%%%%%%%%%%%%%%%%%%%%%%%%%%%%%%%%%%%%%%%%%%%%%%%%%%%%%%%%%%%%%%%%%%

\subsubsection*{Differential operators on $\mathcal M$ associated with the
                $\mathcal H$ action}

The twisted derivations $\partial_x, \partial_y$ considered previously
constitute a $q$-analogue of the notion of vector fields. Their powers
build up arbitrary differential operators. Elements of $\mathcal H$ act
also like powers of generalized vector fields (consider, for instance, the
left action generated by $X_\pm^L,K^L$). Of course, they are differential
operators of a special kind. One can say that elements of $\mathcal H$ act
on $\mathcal M$ as {\sl invariant\/} differential operators since they are
associated with the action of a (quantum) group on a (quantum) space.

A priori, the generators $X_\pm^L,K^L$ can be written in terms of
$x,y, \partial_x, \partial_y$. The coefficients of such combination
can be determined by imposing that equations (\ref{H-products}),
(\ref{actionofHonM}) are satisfied. A rather cumbersome calculation
leads to a unique solution (\cf \cite{CoGaTr-E}) that can be
written simply in terms of the scaling operators \cite{Ogievetsky-2}\
$\mu_x \equiv \one + (q^2-1)(x \partial_x + y \partial_y)$\ , \
$\mu_y \equiv \one + (q^2-1)(y \partial_y)$:

\begin{eqnarray*}
   K^L_- &=& \mu_x \mu_y                 \\
   K^L   &=& \mu_x \mu_x \mu_y \mu_y       \\
   X^L_+ &=& q^{-1} x \partial_y \mu_y \mu_y \\
   X^L_- &=& q y \partial_x \mu_x
\end{eqnarray*}

Notice that elements of $\mathcal M$ acting by multiplication on itself can
be considered as differential operators of order zero. It makes therefore
perfect sense to study the commutation relations between the generators
$x,y$ of the quantum plane and $X_{\pm},K$. This is also done in
\cite{CoGaTr-E}.

%%%%%%%%%%%%%%%%%%%%%%%%%%%%%%%%%%%%%%%%%%%%%%%%%%%%%%%%%%%%%%%%%%%%%%%%%%

\subsection{Star operations on the differential calculus
            $\Omega_{WZ}({\mathcal M})$}
\label{subsec:*-on-Omega}

Given a $*$ operation on the algebra $\mathcal M$, we want to extend it
to the differential algebra $\Omega_{WZ}({\mathcal M})$. This can be
done in two ways, either by using a usual star operation, or by using
a superstar operation (see subsection superstar). Here we use
the ``usual'' star operation formalism, so that the star has to be
involutive, complex sesquilinear, and anti-multiplicative for the algebra
structure in $\Omega_{WZ}({\mathcal M})$. We impose moreover that it
should be compatible with the coaction of $\mathcal F$. The quantum
group covariance condition is, again, just the commutativity of
the $*$, $\Delta_{R,L}$ diagram, or, algebraically,
\begin{equation}
   (\Delta_{R,L} \omega)^* = \Delta_{R,L} (\omega^*) \ .
\label{*-coaction-on-omega-condition}
\end{equation}
However, there is no reason {\it a priori\/} to impose that $*$ should
commute with $d$. In any case, it is enough to determine the action of $*$
on the generators $dx$ and $dy$, since we already determined the $*$
operation on $\mathcal M$ ($* x = x$, $* y = y$).

Taking $\Delta_{R,L} dx = a \otimes dx + b \otimes dy$, we get
$(\Delta_{R,L} dx)^* = a \otimes dx^* + b \otimes dy^*$, to be compared
with $\Delta_{R,L} (dx^*)$. Expanding $dx^*$ as a generic element of
$\Omega^1_{WZ}({\mathcal M})$ (we want a grade-preserving $*$), it can be
seen that the solution $dx^* = dx$ is the only possible one, up to complex
phases. Doing the same with $dy$ we get:
\begin{equation}
   dx^* = dx \ , \qquad dy^* = dy \ .
\label{star-in-Omega}
\end{equation}

\noindent The star being now defined on $\Omega_{WZ}^0 = {\mathcal M}$ and
on the $d$ of the generators of ${\mathcal M}$, it is extended to the whole
of the differential algebra $\Omega_{WZ}$ by imposing the
anti-multiplicative property
$*(\omega_1 \omega_2) = (* \omega_2) (* \omega_1)$.
With this result, it can be checked that
\begin{equation}
   d * \omega = (-1)^p * d \omega \quad \text{when} \quad
                \omega \in \Omega_{WZ}^p \ .
\label{d-*-relation}
\end{equation}

The above involution is not the only one that one can define on the
Wess-Zumino complex. However, any other involution would not be compatible
with the coaction of $\mathcal F$. Loosing the compatibility with the
quantum group is clearly unacceptable, since the main interest of the
Wess-Zumino differential complex rests on the fact that it is compatible
with the coaction.

%%%%%%%%%%%%%%%%%%%%%%%%%%%%%%%%%%%%%%%%%%%%%%%%%%%%%%%%%%%%%%%%%%%%%%%%%%

\section{Non commutative generalized connections on $\mathcal M$ and their
         curvature}
\label{sec:connections}

%%%%%%%%%%%%%%%%%%%%%%%%%%%%%%%%%%%%%%%%%%%%%%%%%%%%%%%%%%%%%%%%%%%%%%%%%%

Let $\Omega$ be a differential calculus over a unital associative algebra
$\mathcal M$, \ie a graded differential algebra with
$\Omega^0 = \mathcal M$. Let $\mathcal M$ be a right module over
$\mathcal M$. A covariant differential $\nabla$ on $\mathcal M$ is a map
${\mathcal M} \otimes_{\mathcal M} \Omega^p \mapsto
  {\mathcal M} \otimes _{\mathcal M} \Omega^{p+1}$, such that
$$
  \nabla( \psi \lambda) = (\nabla \psi) \lambda + (-1)^s \psi \, d \lambda
$$
whenever $\psi \in {\mathcal M} \otimes_{\mathcal M} \Omega^s$
and $\lambda \in \Omega^t$. $\nabla$ is clearly not linear with respect to
the algebra $\mathcal M$ but it is easy to check that the curvature
$\nabla^2$ is a linear operator with respect to $\mathcal M$.

In the particular case where the module $\mathcal M$ is taken as the
algebra $\mathcal M$ itself, any one-form $\omega$ (any element of
$\Omega^1$) defines a covariant differential. One sets simply
$\nabla \one = \omega$, where $\one$ is the unit of the algebra
$\mathcal M$ and we call curvature the quantity $\rho = \nabla^{2}\one$,
$$
   \rho \doteq \nabla \omega = \nabla \one \omega = (\nabla \one) \omega +
                               \one d \omega = d \omega + \omega^2 \ .
$$

%%%%%%%%%%%%%%%%%%%%%%%%%%%%%%%%%%%%%%%%%%%%%%%%%%%%%%%%%%%%%%%%%%%%%%%%%%

\subsection{Connections on $\mathcal M$ and their curvature}

We now return to the specific case where $\mathcal M$ is the algebra of
functions over the quantum plane at a $N$-th root of unity.

The most general connection is defined by an element $\phi$ of
$\Omega_{WZ}^1(\mathcal M)$. Since we have a quantum group action of
$\mathcal H$ on $\Omega_{WZ}$, it is convenient to decompose $\phi$ into
representations of this algebra as obtained in
Section~\ref{subsec:H-action-on-Omega}.
The exact expression of the curvature $\rho = d \phi + \phi \phi$ is not
very illuminating but it can be simplified in several particular
cases (see \cite{CoGaTr-E}).

As we know, the only Hopf star operation compatible with the quantum group
action of $\mathcal H$ on the differential algebra $\Omega_{WZ}$, when $q^N
= 1$, is the one described in Section~\ref{subsec:*-on-Omega} ($dx^* = dx$,
$dy^* = dy$). Imposing the hermiticity property $\phi = \phi^*$ on the
connection leads to constraints on the coefficients. Again we refer to
\cite{CoGaTr-E} for a discussion of the results.

%%%%%%%%%%%%%%%%%%%%%%%%%%%%%%%%%%%%%%%%%%%%%%%%%%%%%%%%%%%%%%%%%%%%%%%%%%

\section{Incorporation of Space-Time}
\label{sec:space-time}

%%%%%%%%%%%%%%%%%%%%%%%%%%%%%%%%%%%%%%%%%%%%%%%%%%%%%%%%%%%%%%%%%%%%%%%%%%

\subsection{Algebras of differential forms over $C^\infty(M) \otimes
   {\mathcal M}$}

Let $\Lambda$ be the algebra of usual differential forms over a space-time
manifold $M$ (the De Rham complex) and
$\Omega_{WZ} \doteq \Omega_{WZ}({\mathcal M})$
the differential algebra over the reduced quantum plane introduced in
Section~\ref{sec:diff-calculus}. Remember that
$\Omega_{WZ}^0 = {\mathcal M}$,
$\Omega_{WZ}^1 = {\mathcal M} \: dx + {\mathcal M} \: dy$, and
$\Omega_{WZ}^2 = {\mathcal M} \: dx \, dy$.
We call $\Xi$ the graded tensor product of these two differential algebras:
$$
   \Xi \doteq \Lambda \otimes \Omega_{WZ}
$$

\begin{itemize}
\item
   A generic element of $\Xi^0 = \Lambda^0 \otimes \Omega_{WZ}^0$ is a
   $3 \times 3$ matrix with elements in $C^\infty(M)$. It can be thought
   as a scalar field valued in $M_3(\CC)$.

\item
   A generic element of
   $\Xi^1 = \Lambda^0 \otimes \Omega_{WZ}^1 \oplus
            \Lambda^1 \otimes \Omega_{WZ}^0$
   is given by a triplet $\omega = ( A_\mu, \phi_x, \phi_y )$, where
   $A_\mu$ determines a one-form (a vector field) on the manifold $M$
   with values in $M_3(\CC)$ (that we can consider as the Lie algebra of the
   Lie group $GL(3,\CC)$), and where $\phi_x, \phi_y$ are $M_3(\CC)$-valued
   scalar fields. Indeed
   $\phi_x (x^{\mu}) \: dx + \phi_y (x^{\mu}) \: dy
      \in \Lambda^0 \otimes \Omega_{WZ}^1$.

\item
   A generic element of
   $\Xi^2 = \Lambda^0 \otimes \Omega_{WZ}^2 \oplus
            \Lambda^1 \otimes \Omega_{WZ}^1 \oplus
            \Lambda^2 \otimes \Omega_{WZ}^0$

   consists of
   \begin{itemize}
   \item
      a matrix-valued $2$-form $F_{\mu \nu} dx^\mu dx^\nu$ on the
      manifold $M$, \ie an element of $\Lambda^2 \otimes \Omega_{WZ}^0$

   \item
      a matrix-valued scalar field on $M$, \ie an element of
      $\Lambda^0 \otimes \Omega_{WZ}^2$

   \item
      two matrix-valued vector fields on $M$, \ie an element of
      $\Lambda^1 \otimes \Omega_{WZ}^1$
   \end{itemize}
\end{itemize}

The algebra $\Xi$ is endowed with a differential (of square zero, of course,
and obeying the Leibniz rule) defined by
$d \doteq d \otimes \id \pm \id \otimes d$. Here $\pm$ is the (differential)
parity of the first factor of the tensor product upon which $d$ is applied,
and the two $d$'s appearing on the right hand side are the usual De Rham
differential on antisymmetric tensor fields and the differential of the
reduced Wess-Zumino complex, respectively.

If $G$ is a Lie group acting on the manifold $M$, it acts also (by
pull-back) on the functions on $M$ and, more generally, on the differential
algebra $\Lambda$. For instance, we may assume that $M$ is Minkowski space
and $G$ is the Lorentz group. The Lie algebra of $G$ and its enveloping
algebra $\mathcal U$ also act on $\Lambda$, by differential operators.
Intuitively, elements of $\Xi$ have an ``external'' part (\ie functions on
$M$) on which $\mathcal U$ act, and an ``internal'' part (\ie elements
belonging to $\mathcal M$) on which $\mathcal H$ acts. We saw that
$\mathcal H$ is a Hopf algebra (neither commutative nor cocommutative)
whereas $\mathcal U$, as it is well known, is a non commutative but
cocommutative Hopf algebra. To conclude, we have an action of the Hopf
algebra ${\mathcal U} \otimes {\mathcal H}$ on the differential algebra
$\Xi$.

%%%%%%%%%%%%%%%%%%%%%%%%%%%%%%%%%%%%%%%%%%%%%%%%%%%%%%%%%%%%%%%%%%%%%%%%%%

\subsection{Generalized gauge fields}
\label{subsec:generalized-gauge-fields}

Since we have a differential algebra $\Xi$ associated with the associative
algebra $C^\infty(M) \otimes {\mathcal M}$, we can define, as usual,
``abelian''-like connections by choosing a module which is equal to the
associative algebra itself. A Yang-Mills potential $\omega$ is an arbitrary
element of $\Xi^1$ and the corresponding curvature, $d \omega + \omega^2$,
is an element of $\Xi^2$. Using the results of the previous subsection, we
see that $\omega = (A_\mu, \phi_x, \phi_y)$ consists of a usual Yang-Mills
field $A_\mu$ valued in $M_3(\CC)$ and a pair $\phi_x, \phi_y$ of scalar
fields also valued in the space of $3 \times 3$ matrices. We have
$\omega = A + \phi$, where $A = A_\mu dx^\mu$ and
$\phi = \phi_x dx + \phi_y dy \in \Lambda^0 \otimes \Omega_{WZ}^1
    \subset \Xi^1$.
We can also decompose $A = A^\alpha \lambda_\alpha$, with $\lambda_\alpha$
denoting the usual Gell-Mann matrices (together with the unit matrix) and
$A^\alpha$ a set of complex valued one-forms on the manifold $M$. Let us
call $\delta$ the differential on $\Xi$, $\underline{d}$ the differential
on $\Lambda$ and $d$ the differential on $\Omega_{WZ}$ (as before).
The curvature is then $\delta \omega + \omega^2$. Explicitly,
$$
\delta A = (\underline{d} A^\alpha) \lambda_\alpha -
           A^\alpha d\lambda_\alpha
$$
and
$$
\delta\phi = (\underline{d}\phi_x) dx + (\underline{d}\phi_y) dy +
             (d \phi_x) dx + (d \phi_y) dy \ .
$$

\noindent It is therefore clear that the corresponding curvature will have
several pieces:
\begin{itemize}

\item
   The Yang-Mills strength $F$ of $A$
   $$
      F \doteq (\underline{d} A^\alpha) \lambda_\alpha + A^2 \quad
               \in \Lambda^2 \otimes \Omega_{WZ}^0
   $$

\item
   A kinetic term ${\mathcal D \phi}$ for the scalars, consisting of three
   parts: a purely derivative term, a covariant coupling to the gauge
   field and a mass term for the Yang-Mills field (linear in the $A_\mu$'s).
   $$
      {\mathcal D \phi} \doteq (\underline{d}\phi_x) dx +
                               (\underline{d}\phi_y) dy +
                      A \phi + A^\alpha d\lambda_\alpha
         \quad \in \Lambda^1 \otimes \Omega_{WZ}^1
   $$

\item
   Finally, a self interaction term for the scalars
   $$
      (d \phi_x) dx + (d \phi_y) dy + \phi^2 \quad
         \in \Lambda^0 \otimes \Omega_{WZ}^2
   $$
\end{itemize}

\noindent Hence we recover the usual ingredients of a Yang-Mills-Higgs model
(the mass term for the gauge field, linear in $A$, is usually obtained
from the ``$A \phi$ interaction'' after shifting $\phi$ by a constant).

By choosing an appropriate scalar product on the space $\Xi^2$, one obtains a
quantity that is quadratic in the curvatures (quartic in the fields) and
that could be a candidate for the Lagrangian of a theory of Yang-Mills-Higgs
type. However, if we do not make specific choices for the connection
(for instance by imposing reality constraints or by selecting one or
another representation of $\mathcal H$), the results are a bit too general
and, in any case, difficult to interpret physically. Actually, many choices
are possible and we do not know, at the present level of our analysis,
which kind of constraint could give rise to interesting physics.

%%%%%%%%%%%%%%%%%%%%%%%%%%%%%%%%%%%%%%%%%%%%%%%%%%%%%%%%%%%%%%%%%%%%%%%%%%
%%%%%%%%%%%%%%%%%%%%%%%%%%%%%%%%%%%%%%%%%%%%%%%%%%%%%%%%%%%%%%%%%%%%%%%%%%

\section{Concluding remarks}

Physical models of the gauge type will involve the consideration of
one-forms. If we restrict ourselves to the ``internal space'' part of these
one-forms, we have to consider objects of the form
$\Phi = \sum \varphi_i \omega_i $. Here $\{ \omega_i \}$ is a basis of some
non-trivial indecomposable representation of $\mathcal H$ (or any other
non-cocommutative quantum group) on the space of $1$-forms, and $\varphi_i$
are functions over some space-time manifold. What about the transformation
properties of the fields $\varphi_i$? This is a question of central
importance, since, ultimately, we will integrate out the internal space
(whatever this means), and the only relic of the quantum group action on the
theory will be the transformations of the fields $\varphi_i$'s. There are
several possibilities: one of them, as suggested from the results of
Section~\ref{sec:space-time} is to consider $\mathcal H$ as a discrete
analogue of the Lorentz group (actually, of the enveloping algebra
$\mathcal U$ of its Lie algebra). In such a case, ``geometrical
quantities'', like $\Phi$ should be $\mathcal H$-invariant (and
$\mathcal U$-invariant). This requirement obviously forces the $\varphi_i$
to transform. Another possibility would be to assume that $\Phi$ itself
transforms according to some representation of this quantum group (in the
same spirit one can study, classically, the invariance of particularly
chosen connections under the action of a group acting also on the base
manifold). In any case, the $\varphi_i$ are going to span some non-trivial
representation space of $\mathcal H$.

Usually, the components $\phi_i$ are real (or complex) numbers and are,
therefore, commuting quantities. However, this observation leads to the
following problem: If the components of the fields commute, then we
get $h.(\varphi_i \varphi_j) = h.(\varphi_j \varphi_i)$, for any
$h \in {\mathcal H}$. This would imply (here $\Delta h = h_1 \otimes h_2$)
\begin{eqnarray*}
   (h_1.\varphi_i)(h_2.\varphi_j) &=& (h_1.\varphi_j)(h_2.\varphi_i) \\
      &=& (h_2.\varphi_i)(h_1.\varphi_j) \ .
\end{eqnarray*}
This equality cannot be true in general for a non-cocommutative
coproduct. Hence we should generally have a nonabelian product for the
fields. In our specific case, there is only one abelian $\mathcal H$-module
algebra, the $3_{odd}$ one. Only fields transforming according to this
representation could have an abelian product. However, covariance strongly
restricts the allowable scalar products on each of the representation spaces
(for instance, in the case of $\mathcal H$ we get both indefinite and
degenerate metrics). This fact is particularly important as one should have
a positive definite metric on the physical degrees of freedom. To this end
one should disregard the non-physical (gauge) ones, and look for
representations such that only positive definite states survive. Thus we
see that the selection of the representation space upon which to build the
physical model is not simple.

The fact of having noncommuting fields has a certain resemblance with the
case of supersymmetry. As the superspace algebra is noncommutative, the
scalar superfield must have noncommuting component fields in order to
match its transformation properties. As a consequence, instead of having
---on each space-time point--- just the Grassmann algebra over the complex
numbers, we see the appearance of an enlarged algebra generated by both the
variables and the fields. It is reasonable to expect that the addition of a
non-trivial quantum group as a symmetry of space forces a more
constrained algebra.

We should point out that the above reasoning is very general, and
is independent of the details of the fields. That is, it relies only in the
existence of a non-cocommutative Hopf algebra acting in a nontrivial way on
the fields.

%%%%%%%%%%%%%%%%%%%%%%%%%%%%%%%%%%%%%%%%%%%%%%%%%%%%%%%%%%%%%%%%%%%%%%%%%%
%%%%%%%%%%%%%%%%%%%%%%%%%%%%%%%%%%%%%%%%%%%%%%%%%%%%%%%%%%%%%%%%%%%%%%%%%%

%%%%%%%%%%%%%%%%%%%%%%%%%%%%%%%%%%%%%%%%%%%%%%%%%%%%%%%%%%%%%%%%%%%%%%%%%%

\end{document}